\documentclass[3p,times,procedia]{elsarticle}
\usepackage{ecrc}
\usepackage{amsmath}

\volume{00}

\firstpage{1}

\journalname{Procedia Computer Science}

\runauth{G. Vincenti}

\jid{procs}

\usepackage{amssymb}
\usepackage{lineno}

\usepackage[figuresright]{rotating}
\begin{document}

\begin{frontmatter}



\dochead{The 2nd International Workshop on Data Mining for Decision Making Support \\ (DMDMS 2018)}

\title{Imprecise temporal associations and decision support systems}

\author{Giovanni Vincenti\correspondingauthor}

\address{University of Baltimore, 1420 North Charles Street, Baltimore, Maryland 21201, United States of America}

\begin{abstract}
The quick and pervasive infiltration of decision support systems, artificial intelligence, and data mining in consumer electronics and everyday life in general has been significant in recent years. Fields such as UX have been facilitating the integration of such technologies into software and hardware, but the back-end processing is still based on binary foundations. This article describes an approach to mining for imprecise temporal associations among events in data streams, taking into account the very natural concept of approximation. This type of association analysis is likely to lead to more meaningful and actionable decision support systems.
\end{abstract}


\correspondingauthor{Tel.: +1-410-837-5886 ; fax: +1-410-837-6252}
\email{gvincenti@ubalt.edu}
\end{frontmatter}

%\linenumbers
%\enlargethispage{-7mm}
\section{Introduction}
\label{sec:intro}

As data mining, decision support systems, and artificial intelligence are quickly entering mainstream technologies, we have to account for the human nature that is innate in people. The recent interest in the field of User Experience (UX) is an important demonstration that software and hardware need to be easily and quickly understood by the users\cite{hassenzahl2006user}. As the user interface is only a filter that mediates between a very subjective user and a very matter-of-fact series of algorithms, we also need to identify ways to let the underlying processes understand and account for the very human concept of imprecision.

Keeping the human nature in mind is not just a concept that is applicable to user interfaces, but also to the processes powering decision support systems\cite{barthelemy2002human}. The vague and often imprecise nature of humans is an essential element to take into account also in the back-end of analytical software. In particular, we have to account for the arbitrary nature of linguistic variables and quantifiers\cite{martinez2010computing}, such as "People purchase {\it more} umbrellas when it is {\it very cloudy}," and temporal associations\cite{ale2000approach}, such as "Sales in football jerseys spike {\it in the days following} a team's victory." The need for a more elastic reasoning model is even more significant when we put the two concepts together, for example when we warn a friend that "traffic gets {\it very intense} on highway XYZ {\it shortly after} it starts raining, even if it is just a {\it light} rain."

This paper presents a practical approach to creating an association analysis system that leads to a more meaningful and actionable set of recommendations. The workshop-oriented nature of the venue in which this paper is presented lends a great opportunity to explore the inner-workings of this analysis methodology as well as some of the fundamental concepts on which it is based.

\section{Background Information}
\label{sec:background}

The process of decision-making cannot be isolated from imprecision as well as temporality, given that the main task of decision support systems is to effectively summarize situations that will then drive actions\cite{aronson2005decision}. Even though time is perhaps the most classic univariate dimension associated with action-detection\cite{yamato1992recognizing}, we can utilize utilize different granularities to study how events are related\cite{ghorbani2017methodology}. Combining these concepts with the basic nature of fuzzy set theory applied to association analysis\cite{chen2009mining}, we can build a powerful engine for decision support systems.

\subsection{Association Analysis}
\label{subsec:association}

One of the principal algorithms in data mining scouts for frequent associations in different commercial transactions \cite{agrawal1994fast}. Its name, Market-Basket Analysis, originates from the goal of identifying which products were frequently purchased together at the grocery store by looking at the pool of items scanned at check-out.

Associations generated through this algorithm are in the form of an implication: $A \Rightarrow B$. The premise $A$ represents the itemset that then triggers the purchase of the itemset represented by the conclusion $B$. Both $A$ and $B$ may contain only one item, or may contain several. A sample association is the following: $\{Peanut Butter, Bread\} \Rightarrow \{Jelly\}$.

As different people may purchase different itemsets, and not everyone who purchases peanut butter and bread may also purchase jelly, the associations must have some metric that quantifies their relevance. The two main metrics are support and confidence. A third metric is lift, but we will not address it in this paper.

Support indicates the strength of the association over all of the associations identified in the analysis. The value is calculated in the following manner:

\begin{equation}
\label{eq:simplesupport}
Sup_{(A \Rightarrow B)} = \frac{\sigma(A \Rightarrow B)}{N}
\end{equation}

where $\sigma(A \Rightarrow B)$ denotes the support count, or frequency, of occurrences of $A \Rightarrow B$ through all associations, divided by the number of all associations identified.

The second metric, confidence, instead quantifies the reliability of the association identified over all the other associations that contain the same trigger itemset. This metric is calculated in the following manner:

\begin{equation}
\label{eq:simpleconfidence}
Conf_{(A \Rightarrow B)} = \frac{\sigma(A \Rightarrow B)}{\sigma(A)}
\end{equation}

where the numerator is the count of instances of the association, and the denominator is the total number of associations that included the same premise, for example both Peanut Butter and Bread, whether the customer may have also purchased Jelly or not. Both metrics give a result in the domain $[0, 1]$.

\subsection{Fuzzy Sets}
\label{subsec:fuzzy}

The domain of computing is often associated with precision, and in particular with the true/false dichotomy that is at the core of this field. This concept is easily implemented through hard boundaries and checks, such as whether a number is greater than another or not, and which is at the base of classic, or Cantorian, set theory. The moment we apply computing to real life though, we find the necessity of implementing solutions that account for ambiguity in terms of classification, or at least that account for the possibility of labeling the same event in multiple ways. The concept of Fuzzy Sets extends the classic set theory to account for this uncertainty \cite{zadeh1965fuzzy}.

\begin{figure}[t]\vspace*{4pt}
\centerline{\includegraphics{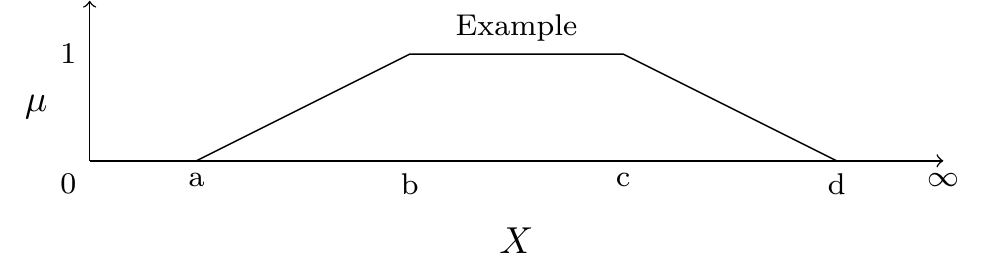}}
\caption{Example of a trapezoidal fuzzy set, also called fuzzy interval.}\vspace*{-6pt}
\label{fig:fuzzyset}
\end{figure}

Although we can have membership functions of several kinds \cite{klir1995fuzzy}, in this work we will only utilize trapezoidal sets, also called fuzzy intervals. Each set is associated with a membership function, as shown in Equation \ref{eq:memfun}.

\begin{equation}
\label{eq:memfun}
\mu_{Example}(x)=
\begin{cases}
0, & x < a\\
\frac{x-a}{b-a}, & a \leq x < b\\
1, & b \leq x \leq c\\
\frac{d-x}{d-c}, & c < x \leq d\\
0, & x > d
\end{cases}
\end{equation}

where the terms $a$, $b$, $c$, and $d$ represent the points that compose a fuzzy interval, shown in Figure \ref{fig:fuzzyset}.

This set will be utilized to assign a linguistic variable \cite{zadeh1975concept} to continuous values. For example, a value of $a-1$ or $d+1$ would not belong to fuzzy set $Example$. Instead values between $b$ and $c$ would have full membership ($\mu = 1$). A value of $x=a+((b-a)/2)$ would have a membership value of $\mu_{Example}(x) = 0.5$. In order to facilitate the discussion, we will refer to $A$ as any instance of $x$ that belongs to stream A, and to the membership weight associated with that linguistic variable as $\mu_{A}$. This will also be done for streams B and C, as well as the temporal relationship $\Delta T$.

\section{Scouting for Imprecise Temporal Associations}
\label{sec:scouting}

The algorithm discussed in this paper addresses two different aspects related to association analysis. First of all, we include a temporal relationship between the triggering events and the consequence, then we account for imprecision using fuzzy sets. The work is formally described by Sudkamp \cite{sudkamp2005examples} and the initial implementation has been published previously by the author and others \cite{vincenti2005data}.

Similarly to association analysis, the associations produced by this method include a premise, which we can call a set of triggers, and a conclusion, which we call a consequence. The triggers itemset currently contains two events, and the consequence itemset contains the temporal relationship as well as an event that is linked to the triggers. The following is an example of an association: $\{ A , B \} \Rightarrow \{ \Delta T , C \}$.

The events $A$, $B$, and $C$ are linguistic variables computed using the process explained in Section \ref{subsec:fuzzy}, and the $\Delta T$ is also a linguistic variable, and it represents the time elapsed between the triggers itemset and the event $C$. A detailed example in Section \ref{subsec:example} will walk through the process of going from a data stream to a set of imprecise temporal associations and the corresponding decision tree.

Since we are working with fuzzy sets, each association cannot be counted as a single instance. The partial membership of any value to more than one set would lead to results with incorrect weights. For this reason, each association will carry a weight calculated in the following manner:

\begin{equation}
\label{eq:assocweight}
W_{\{ A , B \} \Rightarrow \{ \Delta T , C \}} = \mu_A \cdot \mu_B \cdot \mu_{\Delta T} \cdot \mu_C
\end{equation}

Consequently, also the metrics of support and confidence must be adjusted to account for partial memberships of the different events. Support is calculated in the following manner:

\begin{equation}
\label{eq:support}
Sup_{\{ A , B \} \Rightarrow \{ \Delta T , C \}} = \frac{W_{\{ A , B \} \Rightarrow \{ \Delta T , C \}}}{W_{\{ \ast \}}}
\end{equation}

where the numerator is the total weight of all associations $\{ A , B \} \Rightarrow \{ \Delta T , C \}$ divided by the combined weight of all associations, $W_{\{ \ast \}}$. Comparing this way to calculate the support to the traditional method reported in Section \ref{subsec:fuzzy}, we need to note that $N$ may be equal to $W_{\{ \ast \}}$, but it may also be different. This depends on the membership functions that are utilized to classify the stream events, as well as any minimum thresholds that the investigators may set for both support and confidence levels.

Confidence is the second metric, which is calculated using the following equation:

\begin{equation}
\label{eq:confidence}
Conf_{\{ A , B \} \Rightarrow \{ \Delta T , C \}} = \frac{W_{\{ A , B \} \Rightarrow \{ \Delta T , C \}}}{W_{\{ A , B \}}}
\end{equation}

where the numerator is the weight of all associations ${\{ A , B \} \Rightarrow \{ \Delta T , C \}}$ and the denominator is the weight of all associations that have ${\{ A , B \}}$ as the trigger itemset.

\subsection{Example}
\label{subsec:example}

In order to better illustrate the process, we will step through an example that will utilize each element that we discussed so far. We start from a generic set of three streams, as reported in Table \ref{tab:example1stream}.

\begin{table}[h]
\caption{Sample data streams.}
\label{tab:example1stream}
\begin{tabular*}{\hsize}{@{\extracolsep{\fill}}cccc@{}}
\toprule
Timestamp & Stream 1 & Stream 2 & Stream 3 \\
\colrule
0 & 2 & - & - \\
3 & - & 8 & - \\
7 & - & - & 10.5 \\
13 & - & - & 15 \\
1000 & 7 & - & - \\
1003 & - & 2 & - \\
1013 & - & - & 7 \\
\botrule
\end{tabular*}
\end{table}

Each stream event is associated with a timestamp, so that we can establish a temporal relationship between any two events. This will be useful when we are determining whether a pair of triggering events has occurred, and whether a consequence may be linked to a set of triggering events. The streams contain continuous values that are associated with a generic measure of volume, and the timestamp represents a generic unit of time.

In this example, the window between trigger events will be at most of 10 time units, which means that if an event were to occur at Stream 1 at time $T$, we would have a set of triggering events only if an event occurs on Stream 2 between time $T$ and $T+10$. Once the set of triggers is identified, we will evaluate Stream 3 for a consequence. In this case the window between the triggers and the consequence is also of 10 time units.

\begin{table}[h]
\caption{Numerical associations generated from the data stream analysis.}
\label{tab:example1numericalassociations}
\begin{tabular*}{\hsize}{@{\extracolsep{\fill}}ccccc@{}}
\toprule
Association ID & Trigger 1 & Trigger 2 & $\Delta T$ & Consequence \\
\colrule
$num_1$ & 2 & 8 & 4 & 10.5 \\
$num_2$ & 2 & 8 & 10 & 15 \\
$num_3$ & 7 & 2 & 10 & 7 \\
\botrule
\end{tabular*}
\end{table}

The data streams above, in conjunction with the time window constraints, will lead to the identification of three numerical associations, reported in Table \ref{tab:example1numericalassociations}.

\begin{table}[h]
\caption{Boundaries for the fuzzy intervals used to classify the elements of the temporal associations.}
\label{tab:membershipFunctions}
\begin{tabular*}{\hsize}{@{\extracolsep{\fill}}cccccc@{}}
\toprule
Type & Set Name & a & b & c & d \\
\colrule
Data streams & Small Volume & 0 & 0 & 3 & 6 \\
& Medium Volume & 3 & 6 & 9 & 12 \\
& Large Volume & 9 & 12 & 15 & 15 \\
\colrule
Temporal associations & Immediately After & 0 & 0 & 1 & 3 \\
& Short Time After & 1 & 3 & 5 & 7 \\
& Long Time After & 5 & 7 & 10 & 10 \\
\botrule
\end{tabular*}
\end{table}

The next step involves classifying the events represented by the continuous values into linguistic variables. In order to do so, we will have to utilize fuzzy sets. The boundaries of the fuzzy intervals utilized to assign linguistic variables to the stream data as well as the temporal relationships are reported in Table \ref{tab:membershipFunctions}. We are reporting only the values associated with the fuzzy interval, which can be utilized in conjunction with Equation \ref{eq:memfun} to identify the linguistic variable (set name) and to calculate the related membership value.

\begin{table}[h]
\caption{Fuzzy associations generated from the classification of numerical associations through fuzzy sets.}
\label{tab:example1fuzzyassociations}
\begin{tabular*}{\hsize}{@{\extracolsep{\fill}}cccccccc@{}}
\toprule
Association ID & Trigger 1 & Trigger 2 & $\Delta T$ &  Consequence &  Weight &  Support &  Confidence\\
\colrule
$fuz_1$ & Small & Medium & Shortly After & Medium & 0.5 & $0.1\overline{6}$ & $0.25$ \\
$fuz_2$ & Small & Medium & Shortly After & Large & 0.5 & $0.1\overline{6}$ & $0.25$ \\
$fuz_3$ & Small & Medium & Long Time After & Large & 1.0 & $0.\overline{3}$ & $0.5$ \\
$fuz_4$ & Medium & Small & Long Time After & Medium & 1.0 & $0.\overline{3}$ & $1.0$ \\
\botrule
\end{tabular*}
\end{table}

The numeric associations reported in Table \ref{tab:example1numericalassociations} will then be turned into the fuzzy associations reported in Table \ref{tab:example1fuzzyassociations}. One of the first things we can notice is that the number of associations is now greater. This is because not all numeric associations contain values that belong uniquely to one fuzzy set. In the case of association $num_1$ in Table \ref{tab:example1numericalassociations}, we can see that the value associated with Stream 3 is of 10.5, which reports $\mu_{Medium Volume}=0.5$ and $\mu_{Large Volume}=0.5$. This classification into two different sets for the consequence event leads to fuzzy associations $fuz_1$ and $fuz_2$, in Table \ref{tab:example1fuzzyassociations}.

\section{Application to Decision Support Systems}
\label{sec:application}

\begin{figure}[t]\vspace*{4pt}
\centerline{\includegraphics[width=15cm]{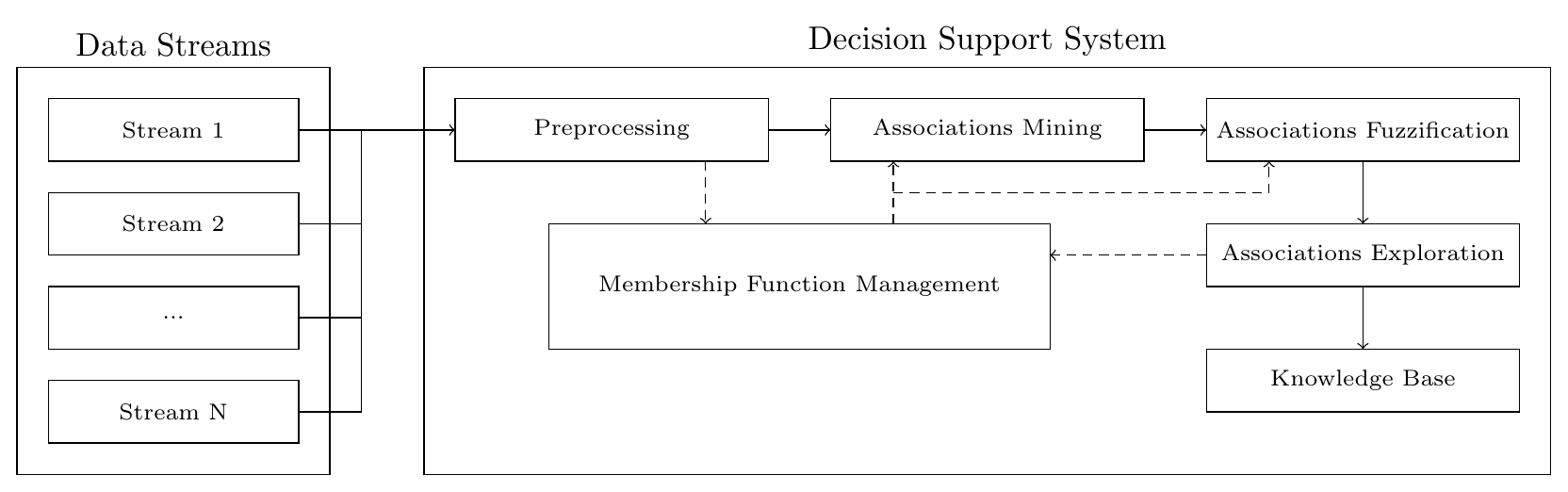}}
\caption{Workflow for the identification of imprecise temporal associations.}
\label{fig:dss}
\vspace*{-6pt}
\end{figure}

The process of mining for imprecise temporal associations is similar to most other data mining processes \cite{kurgan2006survey}. The process utilized by this method is reported in Figure \ref{fig:dss}. The solid lines in the figure represent data flows between processes, while the dashed lines represent influences between processes. The iterative nature of identifying and refining membership functions is the core process that regulates the imprecision associated with any associations discovered. Generating membership functions can be as simple as making educated guesses, or it can be as complex as utilizing elaborate artificially intelligent methods\cite{medasani1998overview}.

\begin{figure}[t]\vspace*{4pt}
\centerline{\includegraphics[width=10cm]{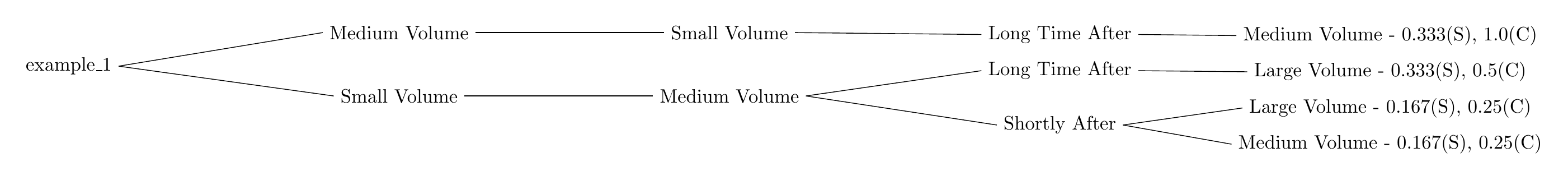}}
\caption{Decision tree generated from the fuzzy associations.}
\label{fig:treeexample1}
\vspace*{-6pt}
\end{figure}

The knowledge base can assume many forms, but the most useful representation that we have found in the context of decision support systems is inspired by decision trees. The associations reported in Table \ref{tab:example1fuzzyassociations} can easily be turned into a decision tree, as reported in Figure \ref{fig:treeexample1}.

This representation leverages a very visual layout that will help decision-makers in the process. The leaf nodes also report Support (Equation \ref{eq:support}) and Confidence (Equation \ref{eq:confidence}) to facilitate the consumption of the findings of this algorithm. The resulting tree can be compared to Frequent-Pattern (FP) Trees \cite{han2000mining}, already in use in data mining \cite{lin2010linguistic}.

\section{Conclusions and Future Work}
\label{sec:conclusions}

The material discussed in this paper guides the reader through the foundations and the basic concepts that are involved with mining for imprecise temporal associations in data streams. The integration of such analysis in the decision-making process is essential, especially given the volume and nature of today's data, as well as its importance in terms of operationalizing the findings quickly and effectively. The work described in this paper has demonstrated its relevance in the fields of medicine and energy utilization, and the findings will be published at a later date.

The next steps for this research focus on algorithmic improvements. In particular, the system is currently limited in the number of streams that it processes to 2 for the triggering itemset and one for the consequence. In order to extend classic association analysis algorithms, this system will also have to accommodate itemsets with one or more streams for either the trigger or the consequence.

Given the nature of continuous stream analysis, the number of potential associations is significant and the runtime for the system can be quite tedious. Incrementing the number of streams to analyze must be accompanied by a different implementation, which should include distributed and parallel processing. Currently the system operates on a single node and does not distributed the work among multiple processors. The distribution of processing during the associations mining phase will significantly increase the capabilities of this system.

Lastly, the process of identifying membership functions is currently driven by domain knowledge and empirical methods. Even though this approach may be appropriate, we will implement other methods of identifying the most meaningful membership functions through statistical methods and the application of artificial intelligence.

%% References

%The citation must be used in following style: \cite{article-minimal} \cite{article-full} \cite{article-crossref} \cite{whole-journal}.

%\bibliography{vincenti_dmdms18}

\begin{thebibliography}{}

\bibitem{agrawal1994fast}
Agrawal R,  Srikant R. Fast algorithms for mining association rules. {\it Proceedings of the 20th International Conference on Very Large Databases (VLDB)} 1994; {\bf 1215}, 487--499.

\bibitem{ale2000approach}
Ale JM, Rossi GH. An approach to discovering temporal association rules. {\it Proceedings of the 2000 ACM Symposium on Applied Computing} (2000); 294--300.

\bibitem{aronson2005decision}
Aronson JE, Liang TP, Turban E. {\it Decision support systems and intelligent systems}. New Jersey: Prentice Hall; 2005.

\bibitem{barthelemy2002human}
Barth{\'e}lemy JP, Bisdorff R, Coppin G. Human centered processes and decision support systems. {\it European Journal of Operational Research} 2002; {\bf 136}(2), 233--252.

\bibitem{chen2009mining}
Chen YL, Weng CH. Mining fuzzy association rules from questionnaire data. {\it Knowledge-Based Systems} 2009; {\bf 22}(1), 46--56.

\bibitem{ghorbani2017methodology}
Ghorbani M, Abessi M. A new methodology for mining frequent itemsets on temporal data. {\it IEEE Transactions on Engineering Management} 2017; {\bf 64}(4), 566--573.

\bibitem{han2000mining}
Han J, Pei J, Yin Y. Mining frequent patterns without candidate generation. {\it ACM SIGMOD Record} 2000; {\bf 29}(2), 1--12.

\bibitem{hassenzahl2006user}
Hassenzahl M, Tractinsky N. User experience-a research agenda. {\it Behaviour \& information technology} 2006; {\bf 25}(2), 91--97.

\bibitem{klir1995fuzzy}
Klir G, Yuan B. {\it Fuzzy sets and fuzzy logic}. New Jersey: Prentice Hall; 1995.

\bibitem{kurgan2006survey}
Kurgan LA, Musilek P. A survey of Knowledge Discovery and Data Mining process models. {\it The Knowledge Engineering Review} 2006; {\bf 21}(1), 1--24.

\bibitem{lin2010linguistic}
Lin CW, Hong TP, Lu WH. Linguistic data mining with fuzzy FP-trees. {\it Expert Systems with Applications} 2010; {\bf 37}(6), 4560--4567.

\bibitem{martinez2010computing}
Mart{\'\i}nez L, Ruan D, Herrera F. Computing with words in decision support systems: an overview on models and applications. {\it International Journal of Computational Intelligence Systems} 2010; {\bf 3}(4), 382--395.

\bibitem{medasani1998overview}
Medasani S, Kim Jm Krishnapuram R. An overview of membership function generation techniques for pattern recognition. {\it International Journal of Approximate Reasoning} 1998; {\bf 19}(3--4), 391--417.

\bibitem{sudkamp2005examples}
Sudkamp T. Examples, counterexamples, and measuring fuzzy associations. {\it Fuzzy Sets and Systems} 2005; {\bf 149}(1), 57--71.

\bibitem{vincenti2005data}
Vincenti G, Hammell RJ, Trajkovski G. Data mining for imprecise temporal associations. {\it Proceedings of the 6th International Conference on Software Engineering, Artificial Intelligence, Networking and Parallel/Distributed Computing} 2005; 76-81.

\bibitem{yamato1992recognizing}
Yamato J, Ohya J, Ishii K. Recognizing human action in time-sequential images using Hidden Markov Model. {\it Proceedings of Computer Vision and Pattern Recognition} 1992; 379--385.

\bibitem{zadeh1965fuzzy}
Zadeh LA. Fuzzy sets. {\it Information and Control} 1965; {\bf 8}(3): 338--353.

\bibitem{zadeh1975concept}
Zadeh LA. The concept of a linguistic variable and its application to approximate reasoning. {\it Information Sciences} 1975; {\bf 8}(3), 199--249.

\end{thebibliography}
%\bibliographystyle{model3a-num-names}



\end{document}